\documentclass[twocolumn,showpacs,preprintnumbers,
               superscriptaddress]{revtex4}
\usepackage{graphicx,epsf}
\newcommand{\bfk}{\mbox{{\boldmath $k$}}}
\newcommand{\bfp}{\mbox{{\boldmath $p$}}}
\newcommand{\bfq}{\mbox{{\boldmath $q$}}}
\newcommand{\bfgamma}{\mbox{{\boldmath $\gamma$}}}

\begin{document}

\title{Pseudogap of Color Superconductivity in Heated Quark Matter}
\author{M. Kitazawa}
\affiliation{Department of Physics,
 Kyoto University, Kyoto 606-8502, Japan}
\author{T. Koide}
\affiliation{Yukawa Institute for Theoretical Physics,
Kyoto University, Kyoto 606-8502, Japan}
\author{T. Kunihiro}
\affiliation{Yukawa Institute for Theoretical Physics,
Kyoto University, Kyoto 606-8502, Japan}
\author{Y. Nemoto}
\affiliation{RIKEN BNL Research Center, BNL, Upton, NY 11973.}

\begin{abstract}
We show that the pseudogap 
of the quark density of states is formed in 
hot quark matter  as 
a precursory phenomenon of the color superconductivity
on the basis of  a low-energy effective theory.
We clarify that the decaying process of
 quarks near Fermi surface to a hole and the diquark soft mode
(qq)$_{\rm soft}$ is responsible 
for the formation of the pseudogap.
Our result suggests that the pseudogap is a universal
phenomenon in strong coupling superconductors.
\end{abstract}

\date{\today}
\maketitle

It is an intriguing problem to examine how rich is 
the phase structure of  the high-density ($\rho$)  QCD 
(Quantum Chromodynamics) matter at vanishing or 
moderate temperature ($T$).
It is now believed  that quark matter at an extremely high density 
undergoes a Cooper instability leading to 
the color superconductivity (CS) 
\cite{ref:Bar,ref:BL,ref:NI,ref:NI2}:
As might be ensured by the asymptotic-free nature of QCD,
the quark matter at extremely high densities is usually 
treated as a Fermi liquid, and 
the mean-field (MF) theoretical approach \`a la BCS theory is
employed;
see, for  reviews on recent exciting development, \cite{ref:review}.

Various complications are, however, expected 
on the nature of quark matter at intermediate baryon densities or 
chemical potential $\mu$ where the strong coupling 
nature of QCD may show up and invalidate the MF approximation \cite{ref:BL}:
The strong coupling may make the so-called Ginzburg region so wide
that precursory fluctuations of the quark-pair field can have a prominent 
strength and may give rise to physically significant effects
above the critical temperature ($T_c$)\cite{ref:KKKN}.
In this Letter, we shall show that the fluctuation effects of the CS
can be so significant that such quark matter 
may share some basic properties with the cuprates
of the high-$T_{c}$ superconductivity (HTSC),
 having a non-Fermi liquid nature above $T_c$.

One of the 
typical non-Fermi liquid properties of the cuprates
is the existence of the {\em pseudogap}, i.e., an anomalous depression of 
density of states (DOS) $N(\omega)$ 
as a function of the Fermion energy $\omega$ 
around the Fermi surface {\em above} $T_c$.
 Although the mechanism of the pseudogap associated with the HTSC is
still controversial, precursory fluctuations of 
the pair field and the quasi-two dimensionality of the
system seem to be basic ingredients to realize the
pseudogap\cite{ref:HTSC,ref:Lok}.
In this work, we shall show that 
 the pseudogap of the quark density of states
exists as a precursory phenomenon of the CS in a considerable
 range of $T$ above $T_c$ even in the three-dimensional system.
We calculate the quark self-energy incorporating the
pre-forming pair-field of the CS for the first time
 in the T-matrix approximation 
(T-approximation\cite{ref:KadBay,ref:HTSC,ref:Schnell}),
where the amplitude fluctuations are
assumed to be dominant over the phase fluctuations of the pair field
\cite{ref:Lok,babaev}. 
Our result shows that the pseudogap may appear also in a relativistic
system solely owing to the amplitude fluctuations, and 
suggests that the pseudogap phenomenon is universal
 in strong coupling
superconductors, irrespective of the dimensionality as suggested from
the study of nuclear matter\cite{ref:Schnell}.
Our results should also provide an insight into
the physics of proto-neutron stars 
and heavy-ion collisions as well, where the $\rho$ is relatively
low and the effect of finite $T$ plays an important role.

To describe a system at relatively low $T$ and $\rho$,
it is appropriate to adopt a low-energy effective theory of QCD 
\cite{ref:NI,ref:BerRaj,ref:SKP}.
Here we employ a simplified version of the instanton-induced 
interaction in the two-flavor case known as
the Nambu-Jona-Lasinio  model\cite{ref:NJL,ref:NJL2}
with the scalar-diquark interaction in the 
chiral limit,
\begin{eqnarray}
\label{eqn:lag}
{\cal L} &=& \bar{\psi}i/\hspace{-2mm}\partial \psi 
+ G_C\sum_{A}(\bar{\psi}i\gamma_{5}\tau_{2}
\lambda_{A}\psi^{C})(\bar{\psi}^{C}i\gamma_{5}\tau_{2}\lambda_{A}\psi) 
\nonumber \\
& & + G_S[(\bar{\psi}\psi)^2+(\bar{\psi}i\gamma_5 \vec{\tau}\psi)^2],
\end{eqnarray}
where $\psi^{C}({\bf x}) \equiv C\bar{\psi}^{T}({\bf x})$, 
with $C = i\gamma_{2}\gamma_{0}$ being the charge conjugation operator.
Here, $\tau_{2}$ and $\lambda_{A}$ mean 
the antisymmetric flavor SU(2) and color SU(3) matrices, respectively.
The coupling $G_S$ and the three dimensional momentum 
cutoff $\Lambda=650$ MeV are determined so as to reproduce the 
pion decay constant $f_{\pi}=93$ MeV and the 
chiral condensate $\langle \bar{\psi}\psi \rangle
=(-250 {\rm MeV})^{3}$ in the chiral limit \cite{ref:SKP}. 
We choose $G_C$ so as to reproduce the phase structure calculated 
using the instanton-induced interaction\cite{ref:BerRaj}, i.e.,
$G_C= 3.11{\rm GeV}^{-2}$ \cite{ref:SKP}.
We neglect the gluon degrees of freedom, especially
their fluctuation, which is known to make the CS phase transition
first order in the weak coupling region\cite{ref:BL,ref:matsuura};
notice that the CS is a type I in this regime.
However, as is emphasized in Ref.~\cite{ref:Pisa}, 
nothing definite is known on the characteristics of the CS
in the
intermediate density region.
In this work, simply assuming that the fluctuations of the
pair field dominates that of the gluon field as is the case 
for type II color superconductors,
we examine the effects of the precursory fluctuations of 
the diquark pair field on the quark sector in the T-approximation.

The DOS $N(\omega)$ is calculated from the 
spectral function ${\cal A}({\bfk},\omega)$ of a single quark,
which is defined through the spectral representation  
of the retarded Green function of the quark field;
\begin{eqnarray}
G^R ({\bfk},\omega)
&=& \int d\omega' \frac{{\cal A}({\bfk},\omega')}{\omega-\omega'+i\eta},
\label{eqn:SR-RGF}
\end{eqnarray}
and accordingly, 
${\cal A}({\bfk},\omega)=-1/\pi \cdot{\rm Im}G^R ({\bfk},\omega)
\equiv -1/\pi \cdot ( G^R - \gamma^0 G^{R\dag} \gamma^0 )/2i $.
From the rotational and parity invariances, 
the spectral function has the following matrix structure:
${\cal A}({\bfk},\omega)=
\rho_{0}({\bfk},\omega) \gamma^0 
- \rho_{\rm v}({\bfk},\omega)\hat{\bfk}\cdot{\bfgamma} +
 \rho_{\rm s}({\bfk},\omega)$,
where $\hat{\bfk} = {\bfk}/|{\bfk}|$ and
 $\rho_{\alpha}$ ($\alpha=0, {\rm v}, {\rm s}$)
still have  color and  flavor indices.
Since the quark number is given by $ N = \int d^3 {\bf x}
\langle \bar{\psi} \gamma^0 \psi \rangle$, 
the DOS is solely given by $\rho_0({\bfk},\omega)$,
\begin{eqnarray}
N(\omega) &=& 4\int \frac{d^3 {\bfk}}{(2\pi)^3}
{\rm Tr}_{\rm c,f}\left[ \rho_{0}({\bfk},\omega) \right],
\label{eq:rho_0}
\end{eqnarray}
with ${\rm Tr}_{\rm c,f}$ 
denoting the trace over color and flavor indices.

The $G^R$ in Eq. (\ref{eqn:SR-RGF}) is given by the analytic continuation
of the imaginary-time (Matsubara) Green function
${\cal G}$, which obeys the following Dyson-Schwinger equation
\begin{eqnarray}
{\cal G}({\bfk}, \omega_n)
= {\cal G}_0({\bfk}, \omega_n)\{1 + 
\tilde{\Sigma}({\bfk},\omega_n){\cal G}({\bfk},\omega_n)\},
\label{eqn:DS-Eq}
\end{eqnarray}
where ${\cal G}_0({\bfk},\omega_n)$ and
$\tilde{\Sigma}({\bfk},\omega_n)$ 
denote the free Green function and the self-energy in the 
imaginary time, respectively. 
In the normal phase, ${\cal G}_0$ is reduced to 
$
{\cal G}_0 ({\bfk},\omega_n)
= [(i\omega_n+\mu)\gamma^0 - \bfk \cdot \bfgamma]^{-1}
$
with the Matsubara frequency $ \omega_n = (2n+1)\pi T $
for fermions.

As was shown in \cite{ref:KKKN},
 the fluctuating diquark pair field
develops a collective mode
(the {\em soft mode} of the CS)
at $T$ above but in the vicinity of $T_c$, in accordance with the
Thouless criterion\cite{ref:Thoul}.
Our point in this work 
is  that the soft mode  in turn
contributes to  the self-energy of the quark field, thereby
it can modify the DOS so much to give rise to a pseudogap.
The  quark self-energy $\tilde{\Sigma}$ owing to the soft mode
 may be  obtained by 
the infinite series of the ring diagrams shown in Fig. 1;
\begin{eqnarray}
\tilde{\Sigma}({\bfk},\omega_n) 
&=& T\sum_{n_{1}}\int \frac{d^3 {\bfk}_{1}}{(2\pi)^3}
\tilde\Xi ({\bfk}+{\bfk}_1, \omega_n+\omega_{n_1}) \nonumber \\
&& \times {\cal G}_0 ({\bfk}_1,\omega_{n_1}), \label{eqn:SE} \\
\tilde{\Xi} ({\bfk},\nu_n)
&=& -8G_C
\left( 1+G_C {\cal Q}({\bfk},\nu_n) \right)^{-1}, \label{eqn:PF}
\end{eqnarray}
with the lowest  particle-particle correlation function 
${\cal Q}({\bfk},\nu_{n})$ \cite{ref:KKKN} and 
 $\nu_{n} = 2n\pi T$ being the Matsubara frequency for bosons.
Notice that the thin quark lines in Fig.1 are the free
Green function, so we have taken the so-called
non-self-consistent approximation, on which we shall make a
comment later.

Inserting Eqs. (\ref{eqn:SE}) and (\ref{eqn:PF}) 
into Eq. (\ref{eqn:DS-Eq}) and performing the analytic continuation 
to the upper half of the complex energy plane, 
we obtain the retarded Green function,
$
G^R(\bfk,\omega) 
= ( G^{-1}_0(\bfk,\omega+i\eta) -\Sigma^R(\bfk,\omega) )^{-1},
$
with 
\begin{widetext}
\begin{eqnarray}
\Sigma^R({\bfp},p^0)
&=& \int \frac{d^3 {\bfq}}{(2\pi)^3}\int \frac{d\omega}{2\pi}
\frac{-{\rm Im}~\Xi^R({\bfp}+{\bfq},\omega)
}{\omega-p^0-E_{\bfq}+\mu-i\eta}
\frac{E_{\bfq}\gamma^0 - {\bfq}\cdot \bfgamma}{2E_{\bfq}}
[
\tanh \frac{E_{\bfq}-\mu}{2T} - \coth\frac{\omega}{2T}
] + (E_{\bfq} \rightarrow -E_{\bfq}), \label{eqn:sigma}
\end{eqnarray}
\end{widetext}
where $E_{\bfq}=\vert \bfq \vert$,
$\Sigma^R({\bfk},\omega) 
= \tilde{\Sigma}(\bfk,\omega_n)|_{i\omega_n=\omega + i\eta}$
and $\Xi^R(\bfk,\omega) 
= \tilde{\Xi}({\bfk},\nu_n)|_{i\nu_{n}=\omega + i\eta}$.
The matrix structure of the self-energy is the same as the spectral 
function,
$\Sigma^R( \bfk ,\omega )
= \Sigma_0( \bfk ,\omega ) \gamma^0 
- \Sigma_{\rm v}( \bfk ,\omega ) \hat{\bfk} \cdot \bfgamma
+\Sigma_{\rm s}( \bfk ,\omega )$
$\equiv \gamma^0(\Sigma_- \Lambda_- + \Sigma_+ \Lambda_+)$,
where $\Lambda_\mp = ( 1 \pm \gamma^0 \bfgamma\cdot\hat{\bfk})/2$
denotes the projection operators onto
the positive and negative energy states,
and accordingly $\Sigma_{\mp}=\Sigma_0\mp\Sigma_{\rm v}$ represents
the self-energies of the particles and anti-particles,
respectively;
notice that $ \Sigma_{\rm s}( \bfk, \omega ) =0 $
in the chirally restored phase
in the chiral limit.
\begin{figure}[tb]
\begin{center}\leavevmode
\epsfxsize=8cm
\epsfbox{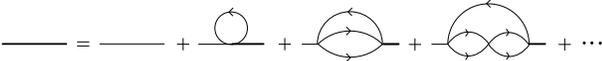} 
\caption{
The Feynman diagrams representing the
quark Green function.
The thin lines represent the free propagator ${\cal G}_0$,
while the bold ones the full propagator ${\cal G}$.
}
\label{fig:self}
\end{center} 
\end{figure}
Then it can be shown that $\rho_0$ is expressed as a sum of the 
positive- and negative-energy parts;
$
\rho_0(\bfk, \omega)=-1/2\pi\cdot\sum_{\alpha=\pm}
{\rm Im} \Sigma_{\alpha}/\{ R_{\alpha}(\bfk, \omega)^2
+({\rm Im}\Sigma_{\alpha})^2\}
$,
where 
$R_{\pm}(\bfk, \omega)=\omega\pm|\bfk|+\mu
-{\rm Re}\Sigma_{\pm}(\bfk, \omega)$.

A remark is in order here:
The Thouless criterion mentioned above
tells us that the denominator of $\Xi^R$, $1 + G_C Q({\bf 0},0)$,
vanishes at $T=T_c$ because of the self-consistency condition for 
the diquark condensate at $T=T_c$\cite{ref:KKKN},
where
$Q({\bfk},\omega) = {\cal Q}({\bfk},\nu_n)|_{i\nu_n 
= \omega + i\eta}$.

For the numerical calculation,
we employ the following cutoff scheme\cite{ref:NJL,ref:KLW}:
First we notice that the imaginary part of $Q$
is free from ultraviolet divergences.
Therefore we first evaluate
the imaginary part exactly without introducing a cutoff,
and then 
use  the dispersion relation with the imaginary part just 
obtained to calculate the real part introducing a 
three dimensional cutoff $\Lambda$.
The imaginary part is nicely found to have 
the following compact form
\begin{eqnarray}
\lefteqn{{\rm Im}~Q ( \bfk ,\omega )
=-N_f (N_c-1)T \frac{(\omega+2\mu)^2 -k^2}{2\pi k} }
\nonumber \\
&\times&\left[
\theta(|\omega+2\mu|-k)\ln\frac{\cosh (\omega+k)/4T}
{\cosh (\omega-k)/4T} \right.
\nonumber \\
&+& \left. \theta(-|\omega+2\mu|+k)\ln\frac{1+e^{-(\omega+k)/2T}}
{1+e^{-(-\omega+k)/2T}}
\right].
\label{eqn:imq}
\end{eqnarray}
We  remark
that each term has the respective interpretation in terms of
the kinetic processes\cite{ref:KKKN}.
We emphasize that 
the compact expression of ${\rm Im}~Q(\bfk,\omega)$
above extremely improves the efficiency of the 
numerical calculations.
Nevertheless the calculations
still involve  multiple integrations for obtaining 
$\Sigma^R$ and hence $\rho_0$.

\begin{figure}[b]
\begin{center}\leavevmode
\epsfxsize=8cm
\epsfbox{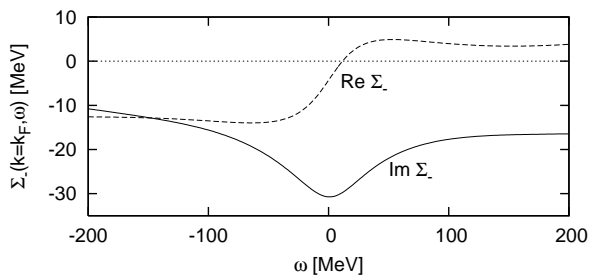} 
\caption{
The self-energy $\Sigma_-$ 
with $ k=k_F$
at $\mu=400$MeV and $ \varepsilon \equiv ( T-T_c )/T_c =0.01 $.
One observes a peak in Im$\Sigma_-$ at $ \omega=0 $
and a rapid increase of Re$\Sigma_-$ at the same $\omega$.
}
\label{fig:sig}
\end{center} 
\end{figure}

\begin{figure}[t]
\begin{center}\leavevmode
\epsfxsize=9.5cm
\epsfbox{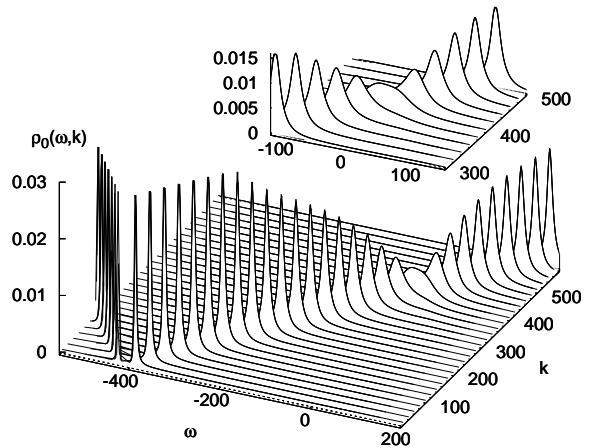} 
\caption{
The spectral function $\rho_0$ at $\mu=400$MeV
and $ \varepsilon=0.01 $.
The upper figure is an enlargement of the one near the Fermi surface.
The peaks at $ \omega=k-\mu $ and $ \omega=-k-\mu$ correspond
to the quark and anti-quark quasiparticles, respectively.
Notice that
there is a depression around $ \omega=0 $,
which is responsible for the pseudogap formation.
}
\label{fig:spc}
\end{center} 
\end{figure}

Since $\rho_0(\bfk, \omega)$ for $\omega >-\mu$
is well approximated solely by the positive-energy part,
we first see the characteristic properties of 
$\Sigma_-$.
Figure \ref{fig:sig} shows  a typical behavior of the 
real and imaginary parts of $\Sigma_-$
with $ k=k_F $ at $ \mu=400 $MeV and the reduced
temperature $ \varepsilon \equiv ( T-T_c )/T_c = 0.01 $
; we remark
that $T_c=40.04$MeV in the present case\cite{ref:KKKN}.
From the figure, one can see that
${\rm Re} \Sigma_-$ shows a rapid increase
around the Fermi energy $ \omega=0$.
The quark  dispersion relation $ \omega=\omega_-(k)$ therefore 
also shows a similar behavior around the Fermi surface;
$\omega_-(k)$ is the solution of
$R_-(\bfk, \omega)=0 $.
Hence the density of states proportional
to $ (\partial \omega_-/\partial k)^{-1}$ 
becomes smaller near the Fermi surface, which suggests the
existence of a pseudogap, 
provided that the imaginary part ${\rm Im}\Sigma_-$ is neglected,
which will be discussed shortly.
One  can also see that  $\omega_-(k= k_F)\simeq k-\mu$
since ${\rm Re} \Sigma_-$ at $ \omega=0 $ is vanishingly small,
which will 
be found important for
the pseudogap formation
around the Fermi surface.

A numerical calculation shows that
as the momentum $k$ is varied from $k_F$,
the peak of $|{\rm Im} \Sigma_-|$
at $ \omega\approx0 $ seen in Fig.\ref{fig:sig} 
moves along $ \omega \approx -k+\mu$\cite{nonfermi}.
This means that the quasiparticles
with  this energy are dumped modes.
Figure \ref{fig:self} tells us that ${\rm Im} \Sigma_-$ describes 
a decay process of a quark to a hole and a diquark, 
q$\to$h$+$(qq), where the hole is on-shell with a free dispersion 
relation $\omega_h=\mu-|\bfk_h|$.
The essential point for the pseudogap formation is that
the above process is  enhanced when the diquark (qq) 
makes a collective mode,
 which we have emphasized is the case;
 the diquark soft mode (qq)$_{\rm soft}$ has a prominent strength at 
vanishingly small
energy $\omega_s$ and momentum $\bfk _s$ near $T_c$.
Owing to the energy-momentum conservation,
the energy-momentum of the decaying particle $(\omega_p, \bfk_p)$
should satisfy
$\omega_p+\omega_h=\omega_s\simeq 0$ and
$\bfk _p+\bfk_h=\bfk _s\simeq 0$.
It means that 
when the decaying particle has almost the same energy as a
free quark has, $|{\rm Im}\Sigma_-(\bfk, \omega)|$ has the largest
value.

The spectral function
 $\rho_0(\bfk, \omega)$ is shown in
Fig.\ref{fig:spc},
at the same $\mu$ and $\varepsilon$ as those in Fig.\ref{fig:sig}.
One can see two families of peaks around 
$ \omega = \omega_-(k) \approx k-\mu $ and $ \omega = -k-\mu $,
which correspond to the quasiparticle peaks of
the quarks and anti-quarks, respectively.
A notable point is that
the quasiparticle peak has a clear depression around $ \omega=0$,
i.e., the Fermi energy.
The mechanism for the depression is easily understood
 in terms of the characteristic properties of the
self-energy 
 mentioned above: In fact, $R_-(k_F, \omega\simeq \mu -k_F)\simeq 0$ and
$|{\rm Im} \Sigma _-(\bfk, \omega)|$ becomes large
when $ \omega \approx -k+\mu$.
Thus 
$\rho_0(k\simeq k_F, \omega\simeq 0)\simeq 
-1/(2\pi{\rm Im} \Sigma_-(k_F, \omega\simeq 0))$,
which is found to be suppressed.

\begin{figure}[t]
\begin{center}\leavevmode
\epsfxsize=8cm
\epsfbox{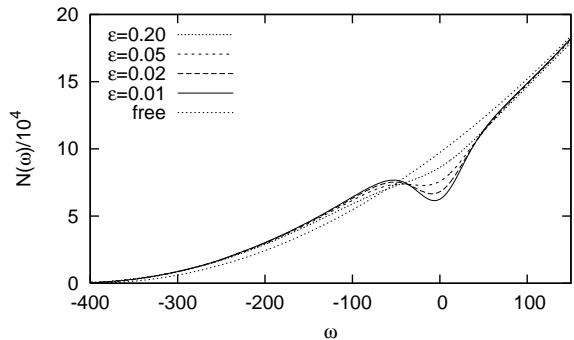} 
\caption{
Density of states at $\mu=400$MeV and various 
$\varepsilon\equiv(T-T_C)/T_C$.
The Dotted line shows that of the free quarks.
A clear pseudogap structure is seen, which survives
up to $ \varepsilon\approx 0.05$.
}
\label{fig:dos}
\end{center} 
\end{figure}

Integrating $\rho_0$, one obtains the DOS $N(\omega)$:
Figure \ref{fig:dos} shows the DOS at $\mu=400$MeV
and various values of the reduced temperature $\varepsilon$
 together with that of the free quark system,
$N_0(\omega)$.
As anticipated,
one can see a remarkable depression of $N(\omega)$,
i.e., the {\em pseudogap}, around 
the Fermi energy $ \omega=0 $;
$N(\omega)/N_0(\omega)|_{\omega=0} \simeq 0.64$
at $ \varepsilon=0.01 $.
One sees that 
the smaller $\varepsilon$, the  more remarkable the 
rate of depression.
The clear pseudogap structure
survives even at $ \varepsilon=0.05$.
One may thus conclude that
there is a  pseudogap region
within the normal phase above $T_c$ up to
$T=(1.05\sim1.1)T_c$ at $\mu=400$MeV, for instance.
A numerical calculation shows that $\varepsilon$ dependence of the width
of the pseudogap region
hardly changes 
for $320$MeV$<\mu<500$MeV.

We notice that the pseudogap region
obtained in the present work is more than one order of magnitude
 wider in the unit of $\varepsilon$
than in the nuclear matter\cite{ref:Schnell}
where the clear pseudogap is seen up to $ \varepsilon\approx0.0025 $.
This is just a reflection of the
strong coupling nature of the QCD at intermediate density region.
Our result obtained for a three-dimensional 
system tells us  that
a considerable pseudogap can be 
formed without the help of
the low-dimensionality as in the HTSC and
that the pseudogap phenomena in general
may be  universal  in any strong coupling
superconductivity.

In this Letter,
we have found that the pseudogap can be formed 
as a precursory phenomenon of the CS
in a rather wide region of $T$ above $T_c$.
This may imply that quark matter 
shares some basic properties with
the cuprates of the  HTSC.
It should be noted that
our work is the first calculation to show the formation
of the pseudogap in the relativistic framework.

In the present work, we have employed the 
non-self-consistent T-approximation, while
the self-consistent approximation apparently seems better.
However, it may not be the  case\cite{Fuj02};
higher-order terms with the
vertex corrections, which are usually discarded in the self-consistent 
approximation, cancel with each
other, which means that the lowest-order calculation 
such as ours is more reliable than
the self-consistent one\cite{Fuj02}.


As a future problem, one should consider how
to observe the pseudogap in heavy ion collisions or
 proto-neutron stars.
It is also interesting 
to see what would happen if the phase transition of the CS
is strong first order.

We thank K. Iida for and D. Rischke for 
their valuable comments and drawing our
attention to \cite{ref:Pisa} and \cite{Boy01}, 
respectively.
T. Kunihiro is supported by
Grant-in-Aide for Scientific Research by
Monbu-Kagaku-sho (No.\ 14540263).


\begin{thebibliography}{99}
\bibitem{nonfermi}
The peak structure of the
 $|{\rm Im} \Sigma_-|$ around $ \omega=0$ is a typical
non-Fermi liquid behavior, although
the wave function renormalization
$Z$ of the quasiparticle at the Fermi surface is found to be
$
  Z  = [ 1+ d{\rm Re\ } \Sigma_-(\bfk, \omega)/d\omega
    ]^{-1}_{k=k_F,\omega=0} \approx 0.7,
$
at $\mu=400$MeV and $\varepsilon=0.01$,
which suggets that the non-Fermi liquid picture is not 
complete; see for other possible origins of the non-Fermi liquid
 behavior of quark matter, \cite{Boy01}.
\end{thebibliography}
\end{document}